\documentclass[prl,aps]{revtex4}
\usepackage{graphicx}
\begin{document}
\title{Looking back at superfluid helium} 
\vskip2in
\author{S\'ebastien Balibar\footnote{e-mail address: 
balibar@lps.ens.fr}}
\vskip1in
\affiliation{Laboratoire de 
Physique Statistique de l'Ecole Normale Sup\' erieure,\\associ\'e au CNRS 
et aux Universit\' es Paris 6 et 7,\\
24 rue Lhomond, 75231 Paris Cedex 05, France.}

\vskip1in

\vskip0.5in
\begin{abstract}
    \baselineskip = 2\baselineskip
A few years after the discovery of Bose Einstein condensation in 
several gases, it is interesting to look back at some properties of 
superfluid helium. After a short historical review, I comment shortly 
on boiling and evaporation, then on the role of 
rotons and vortices in the existence of a critical velocity in 
superfluid helium. I finally discuss the existence 
of a condensate in a liquid with strong interactions,
and the pressure variation of its superfluid transition temperature.

\end{abstract}
\maketitle
\baselineskip = 2\baselineskip
\vskip1in
\section*{The discovery of superfluidity}

In january 1938, J.F. Allen and A.D. Misener~\cite{Allen1} published
the experimental evidence that the hydrodynamics of liquid helium was
not classical below 2.2~K. In the same issue of Nature, Kapitza
introduced the word ``superfluid'' to qualify this anomalous
behavior~\cite{Kapitza}.  Soon after this discovery, F. London
suggested that superfluid helium forms a macroscopic liquid matter
wave, as a consequence of ``Bose-Einstein condensation'' (BEC). 65
years later, it is generally accepted that London was right, but
it happenned to be very difficult to prove.  On the contrary, when BEC
was discovered in cold alkali vapors (in 1995), the experimental
evidence was immediately very clear. The superfluidity of these cold
gases was soon demonstrated as well. It is thus interesting to
look back at some properties of superfluid helium now.  As we shall
see, not everything is completely understood in the properties of
superfluid helium. In this short review, I have selected only a few 
properties of superfluid helium. For a complete description, one can 
look at the historical book written by 
J. Wilks~\cite{Wilks}, or the one by P. Nozi\`eres and D. Pines 
on Bose liquids~\cite{Nozieres}. 
Here, I only wish to recall some aspects of the early history of this
subject in order to show that fundamental questions came up very soon
in this field.  Then, I whish to briefly discuss three particular 
topics:

1- boiling and evaporation because this is what makes superfluid helium
look different from normal helium in a dewar.

2- the mechanisms which determine the ``critical velocity''

3- the existence of a condensate in superfluid helium
and the pressure variation of 
the superfluid transition temperature.

In the years 1928-32 in Leiden,
W.H. Keesom and his co-workers had already realized that liquid helium exhibited
surprising properties below about 2.2~K~\cite{Keesom1927}.  They had
found a peak in the temperature variation of the specific heat at this
temperature.  The shape of this peak resembling the greek letter
lambda, Keesom called the superfluid transition temperature
the lambda point or ``lambda temperature
$T_{\lambda}$''. He also called ``helium I'' the liquid above $T_{\lambda}$
and ``helium II'' the liquid below $T_{\lambda}$.  In his laboratory
in Leiden, Keesom had also discovered that helium II was able to flow
through very tiny pores as soon as in 1930~\cite{Keesom1930}. In 1932
in Toronto, J.C. McLennan, H.D.~Smith and J.O.~Wilhelm then discovered
that liquid helium ceased boiling when cooled down through
$T_{\lambda}$~\cite{McLennan}.  This was soon attributed to an
exceptionnally high thermal conductivity, as evidenced by successive
experiments by W.H.~Keesom again~\cite{Keesom1936} and by J.F.~Allen in
Cambridge~\cite{Allen1937}. J.F. Allen had been hired with R.
Peierls to replace Kapitza there, because Kapitza had been forced by 
Stalin in 1934 to stay in
Moscow and to quit his research position in Cambridge\cite{SFP}. 
Since the high
thermal conductivity was attributed to some kind of convection or 
turbulence, researchers looked
at the flow of liquid helium.  In 1935 in Toronto, J.O. Wilhelm, A.D.
Misener and A.R. Clark had measured the viscosity of liquid helium
with a torsion pendulum and found that it decreased sharply below
$T_{\lambda}$~\cite{Wilhelm}.  When A.D~Misener moved to Cambridge in
order to prepare his doctorate with J.F.~Allen, they started together 
a systematic
study of the flow of liquid helium through capillaries.  Classical
fluids obey the Poiseuille law: the flow rate is proportional to the
pressure difference accross the capillary and to the fourth power of
the capillary radius.  Instead of this, Allen and Misener found that,
below $T_{\lambda}$, the flow rate was not only high, it was 
independent of the pressure and independent of the capillary radius which they
had changed by a factor 50!  Clearly, this liquid was
not classical.  In a sense, this was also what Kapitza claimed in his
article~\cite{Kapitza}, since he introduced the word ``superfluid'' in
connection with the already know ``superconductivity''.  However,
Kapitza did not justify his remarkable intuition that superfluidity
and superconductivity were related phenomena, and his 1938 article 
contained no quantitative measurements. He performed remarkable 
experimental measurements during the following years.

In  february 1938, J.F. Allen and H. Jones published another estonishing 
discovery. They had found a remarkable
thermomechanical effect~\cite{AllenJones}.  When heating superfluid 
helium on one side of
a porous medium or a thin capillary, the pressure increased
sufficiently to produce a little fountain at the end of the tube
which contained the liquid.  The ``fountain effect'' was another spectacular
phenomenon which was impossible to understand within classical
thermodynamics.  This is really what triggered Fritz London's
thinking~\cite{London}. London new that, in this liquid, there were large
quantum fluctuations responsible for the large molar volume, 
and he proposed that the superfluid transition
at $T_{\lambda}$ was a consequence of Bose-Einstein condensation (BEC).  65
years later, the connection between
superfluidity and BEC is still a matter of debate and study. Before 
discussing this (in Section III), I wish to say a few words about 
boiling and evaporation (section I), and , in Section II, I briefly 
review the problem of the critical velocity, in connection with 
Landau's ``two fluid
model''~\cite{Landau}, 
the existence of ``rotons'' and quantized vortices. One of my motivations in 
section II is to allow a comparison 
with observations of a critical velocity and vortices in gaseous 
superfluids.

\section{Boiling, evaporation and cavitation}

\begin{figure}[t]
\centerline{\includegraphics[height=8cm]{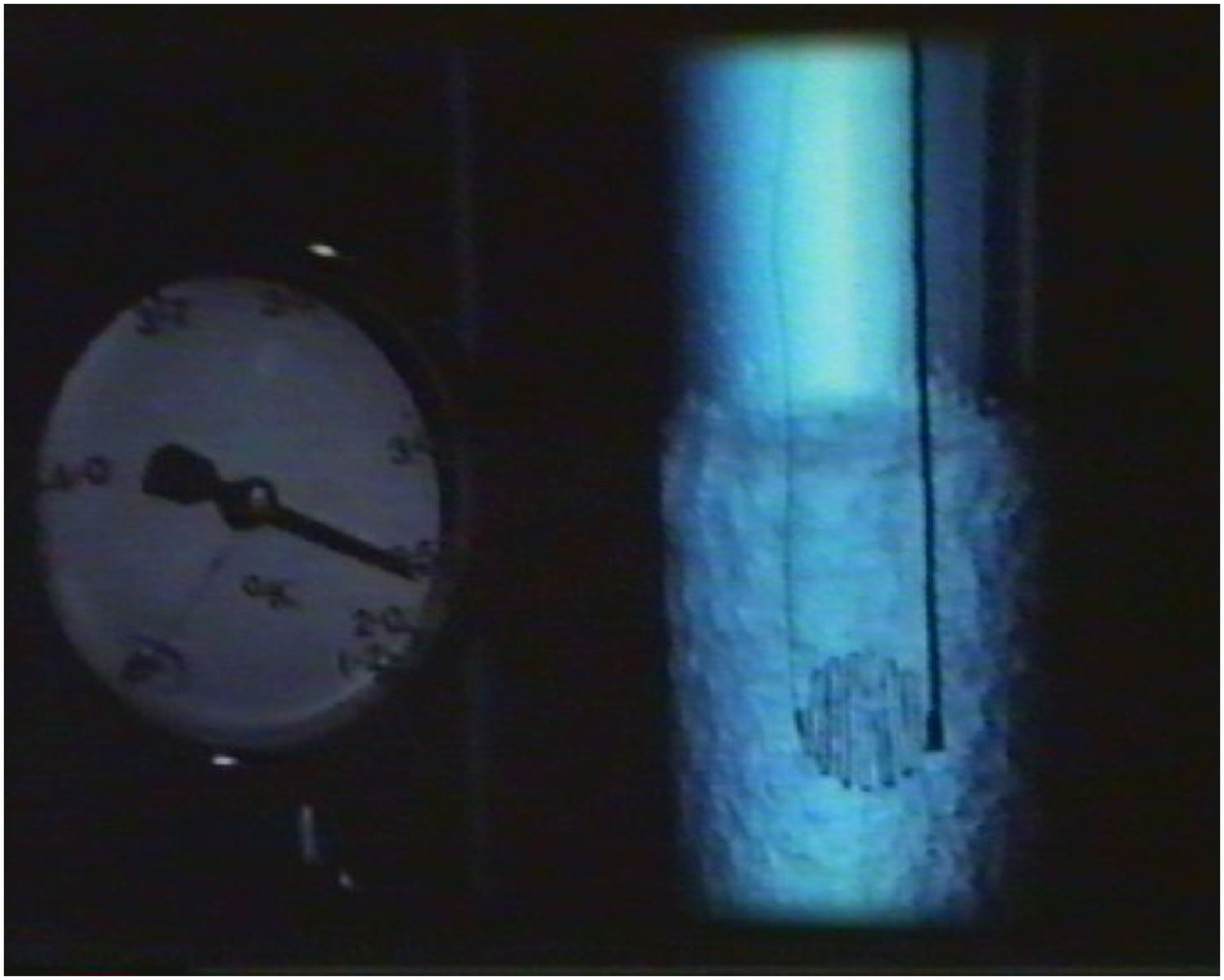}}
\vskip0.5in
\centerline{\includegraphics[height=8cm]{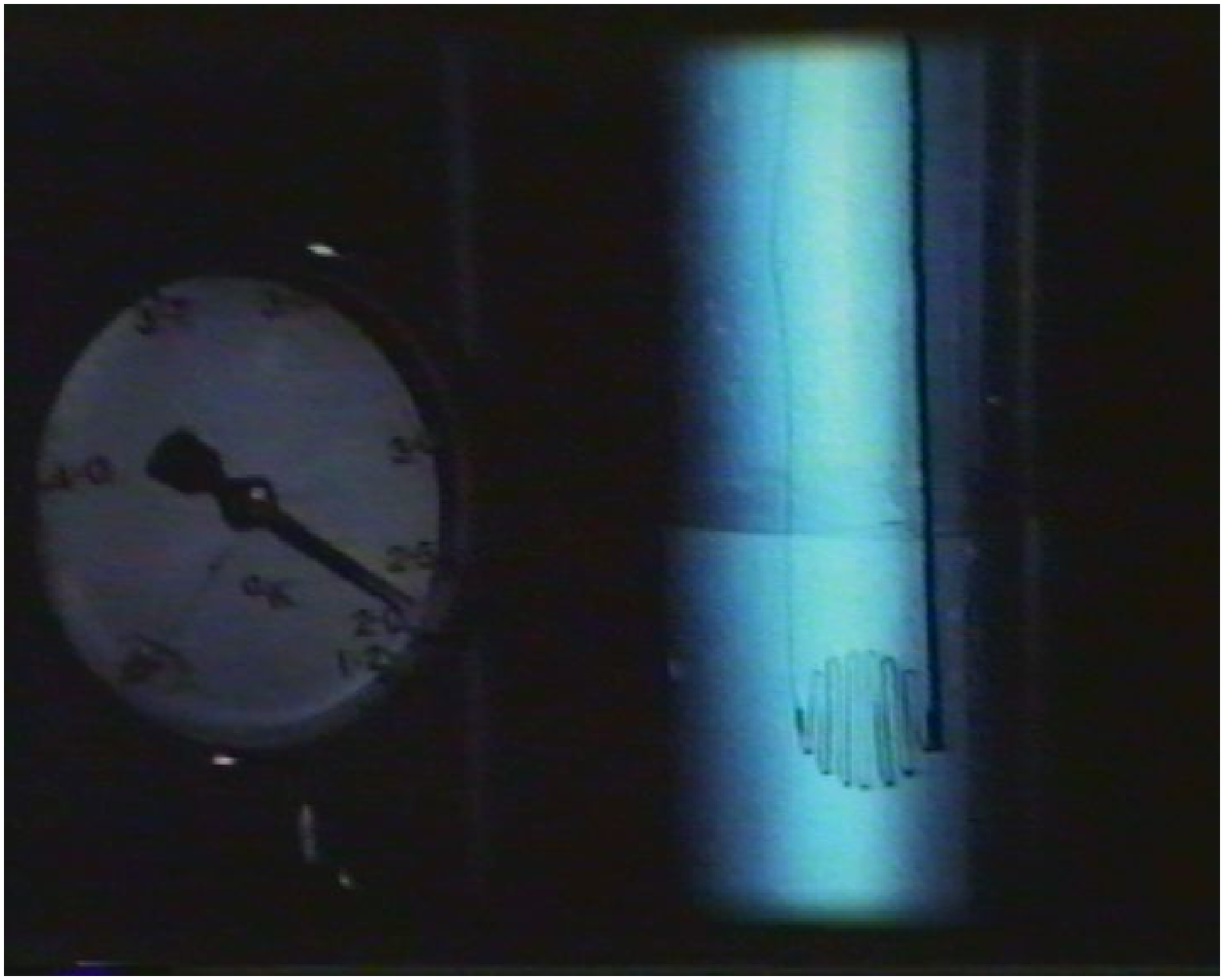}}
\vskip0.5in
\caption{As shown by these two images from a film by J.F. Allen and J. 
Armitage, superfluid helium stops boiling below $T_{\lambda}$. This 
is due to its large thermal conductivity. The top image is taken at 
2.4~K as indicated by the needle of the thermometer on the left. The 
bottom image is taken just below the lambda transition.}
\label{fig:Allen}
\end{figure}

The easiest way to see superfluidity is to look at  
liquid helium in a dewar, while pumping on it so as to cool it below 
$T_{\lambda}$ = 2.2~K (see Fig.~\ref{fig:Allen}). When crossing 
$T_{\lambda}$, the liquid stops boiling. This is because the thermal 
conductivity of liquid helium has suddenly increased, so that the 
temperature is very homogeneous. The walls are no longer warmer than 
the surface. They no longer provide efficient nucleation sites for 
bubbles. As a result, superfluid helium evaporates from its free 
surface, instead of showing bubble nucleation on hot defects. 
Superfluid helium looks quiet, and it is tempting, although not 
rigorous, to take this as an 
illustration of quantum order getting established. At least this shows 
a striking reality: a liquid made of very simple atoms can exist in 
two different states! It happens that I worked on the evaporation of 
superfluid helium~\cite{Balibar78} so that i would like to add a few 
comments on the evaporation of superfluid helium, an interesting 
property which is rarely mentionned. 
We have verified the prediction by P.W. 
Anderson that, at low enough temperature, this evaporation is a 
quantum process which is analogous to the photoelectric 
effect~\cite{Anderson}. A photon incident on a metal surface can 
eject one electron elastically: the electron 
 kinetic energy is the difference between the photon energy and the 
 binding energy of the electron to the metal. Similarly, we have shown 
 that atoms can be ejected from superfluid helium with a kinetic 
 energy which is close to 1.5~K, the difference between the energy of incident 
 rotons (whose minimum is 8.65~K) and the binding energy of 
 atoms to the liquid, which is nothing but the latent heat of 
 evaporation per atom (7.15~K). This phenomenon was later 
 called ``quantum evaporation'' by A.F.G.~Wyatt who made a long 
 quantitative study of it. A few open issues remain in this subject, 
 such as the exact measurement and calculation of evaporation 
 probabilities by phonons or rotons as a function of energy, momentum 
 and incidence angle, or the nature of possible inelastic processes as 
 well.
 
 Furthermore, one should not believe that boiling, more generally 
 cavitation or bubble nucleation,  is impossible 
 in superfluid helium.  On the contrary, this is another interesting 
 subject on which we also worked~\cite{Balibar02}.
 A strong local heat input can generate 
 bubbles in superfluid helium, of course. Moreover, one can apply a large 
 stress, that is a large negative pressure to liquid helium at low 
 temperature and also look for the nucleation of bubbles. This is a way to 
 test the internal cohesion of this liquid. We have observed 
 this phenomenon near -9.5 bar, the stability limit of 
 liquid helium under stress, also called the ``liquid-gas spinodal 
 limit''. It allowed us also to obtain some information on 
 superfluidity at reduced density as we shall see briefly in Section 
 III.

\section{The critical velocity: rotons and vortices}

The vanishing of boiling is spectacular but the existence of flow 
without dissipation is of course the more fundamental property which 
signals superfluidity. In may 1938, one month only after London had 
published his suggestion of a Bose-Einstein condensation, L. Tisza 
introduced a ``two fluid model'' which qualitatively explained the 
observations. He considered that ``the atoms belonging to the lowest 
energy state do not take part in the dissipation. Thus the viscosity 
of the system is entirely due to the atoms in the excited states''. 
He understood that this liquid could be considered as a mixture of two 
components: a superfluid component which carries no entropy and has zero 
viscosity, plus a normal component, made of atoms in excited states, 
which is viscous and conducts heat. The ratio of one to the other 
component was determined by the temperature: at zero temperature 
there is only the superfluid and at the lambda transition there is 
only the normal component. The important point was to understand that 
these two components could move independently. This very pure liquid had two 
independent velocity fields, something London could not easily admit.
From these considerations Tisza predicted that when superfluid helium 
flowed through a capillary, the normal component, being viscous, 
was blocked while the superfluid component could flow without 
dissipation. As a consequence, viscosity measurements could not give the 
same result when done in an open geometry, as in the torsion 
pendulum used in Toronto, or in a restricted geometry like a thin 
capillary or a porous medium. This resolved the apparent discrepancy 
between different experiments at that time. Moreover, he made an even more 
surprising prediction, namely that ``a temperature gradient should 
arise during the flow of helium II through a thin capillary''. Tisza 
was right! Helium cools down when flowing so. Furthermore, he 
explained the fountain effect as the reverse of the previous one: the 
superfluid component moves in the direction of the hot side where the 
pressure increases. His 
1938 article was only one column long. He developped his ``two fluid 
model'' in two subsequent articles~\cite{Tisza2} where he made another fundamental 
prediction: thermal waves should exist where the superfluid and 
normal components move out of phase.

Nearly all the predictions by Tisza were proved correct by subsequent 
experiments. However, his model appeared slightly incorrect because 
the superfluid and normal components could not be defined in relation
to a Bose gas. The 
two fluid model was subsequently developped by Landau in his 1941 
article, with no reference, this time, to BEC. He considered energy states 
of the liquid instead of individual atoms. He attributed a mass 
density $\rho_{s}$ to the part of the fluid which occupies the ground 
state. The rest of the density $\rho_{n} = \rho - \rho_{s}$ 
corresponded to the excited states. Furthermore, he had a remarkable 
intuition about the nature of these excited states. He assumed that single 
particle states were replaced by 
 collective modes of two different kinds: phonons, i.e. sound 
 quanta, and quantized vortices which he called ``rotons''.
 Phonons had basically a linear 
 dispersion relation $\omega = ck$ and rotons had a quadratic one with 
 a gap ($\epsilon = \Delta + p^{2}/2\mu$), so that no excitation had a 
 phase velocity $\epsilon/p$ less than a certain minimum value $v_{c}$.
 From this 
 assumption, Landau calculated the thermodynamics of helium II, the 
 magnitude of thermal waves called ``second sound'', etc. The major 
 breakthrough in this article is the introduction of elementary 
 excitations as quantized collective modes which led him to the 
 crucial prediction of a critical velocity $v_{c}$. He 
 explained that, due to the necessary conservation of energy and 
 momentum, the superfluid component in motion at a velocity $v$ could 
not slow down by exciting collective modes if $v$ was less than 
$v_{c}$. In the absence of rotons, $v_{c}$ would have been the sound 
velocity c. Landau modified the dispersion relation of rotons in his 
1947 article~\cite{Landau47} in order to obtain a better agreement 
with thermodynamic properties. The phonon branch evolved continuously 
into a roton branch $\epsilon = \Delta + (p-p_{0})^{2}/2\mu$ around the finite
momentum $p_{0}$. Given this phonon-roton spectrum, Landau's critical velocity 
was predicted to be about 60 m/s at low pressure.

\begin{figure}
\centerline{\includegraphics[width=0.6\linewidth]{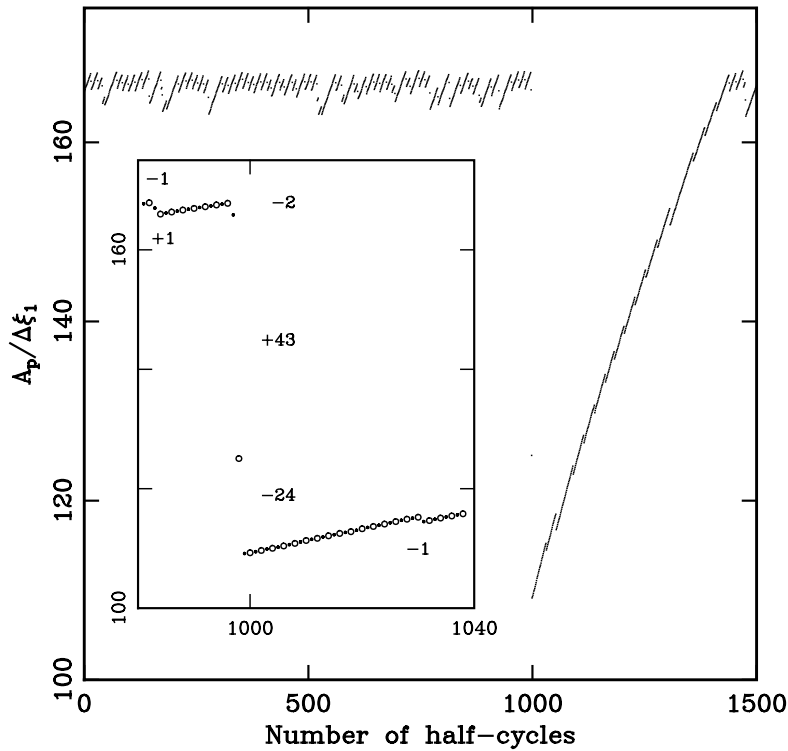}}
\caption{The quantized dissipation in the experiment by O. Avenel and E. 
Varoquaux. The vertical coordinate is the amplitude of oscillation of 
a superfluid flow through a small aperture. It is proportional to an 
average  velocity in the hole. The horizontal coordinate is 
proportional to time. The velocity changes by finite jumps which are 
quantized. Single jumps correspond to one quantized vortex line 
crossing the hole. Multiple jumps are also observed.}
 \label{fig:Varoq}
 \end{figure}

M. Cohen and R.P. Feynman~\cite{Cohen} then proposed to verify
Landau's dispersion relation in neutron scattering experiments. This 
was done with great accuracy by several groups, in particular by Henshaw and 
Woods~\cite{Henshaw}. Of course, it was a strong support to Landau's theory. In 
the mean time, superfluidity was proved not to exist in liquid helium 
3 at similar temperatures, and this was a strong support to London 
and Tisza. As for the existence of Landau's critical velocity, things 
appeared difficult. In fact, Landau had proposed this mechanism 
but he had not ruled out the possibility that other mechanisms could exist at 
lower velocity. The existence of quantized vortex lines was proposed 
in 1949 by Onsager~\cite{Onsager} who predicted that the circulation of their 
velocity should be quantized as multiples of $h/m$. Independently in 1955,
Feynman~\cite{Feynman} explained that such
vortices should lead to a much lower critical velocity of order 
$v_{c}= (\hbar / md) \ln (d/a)$ where d is the size of the 
container (a vessel or a capillary diameter) and $a$ the atomic size 
of the vortex core. Feynman was right. The most beautiful 
evidence for the vortex mechanism was obtained by  
Avenel and Varoquaux~\cite{Varoquaux} who could see quantized
dissipation events in a flow through a small aperture
(see Fig.~\ref{fig:Varoq}). Moreover, their analysis of vortex 
nucleation shows a remarkable crossover from thermally activated 
nucleation above a crossover temperature of order 0.2~K to nucleation 
by quantum tunneling in the low temperature limit. A similar crossover 
was found in studies of superconducting Josephson  junctions, also in 
our studies of cavitation in superfluid helium 4.

 As for Landau's critical velocity, it was also observed, but in  
 cases where the system size was very small only. P. McClintock et al. 
 measured the mobility of electrons moving in superfluid helium under 
 special conditions. These electrons repell the electronic clouds of 
 neighbouring atoms so as to form little empty cavities around them, 
 whose typical radius is 10 \AA. At low pressure and small electric field,  
 these electron bubbles trap vortex rings as was checked from by their energy vs 
 momentum curve. At high pressure (25 bar, near the melting pressure) 
 and high electric field (from 20 to 2000 V/cm), the electron bubbles could in 
 fact be accelerated to velocities slightly larger then Landau's 
 prediction. This was understood by Bowley and Sheard as a consequence 
 of rotons being emitted in pairs~\cite{McC}. This whole subject would 
 need a much longer review, but the comparison with gaseous 
 superfluids is interesting. Superfluid gases have no rotons in their 
excitations. Landau's critical velocity is thus associated with 
phonons. It is observed if the moving object is small (an atom), but 
if it is larger (a laser beam for exemple), then dissipation occurs at 
lower velocity. Just as in superfluid helium the nucleation of vortices 
is invoked to understand the low value of the critical velocity in 
this case. The comparison between quantized vortices in superfluid 
helium and in superfluid gases has been beautifully extended to
the observation of 
 similar vortex arrays in both cases (see J.~Dalibard and C.~Salomon, this 
conference). 

\section{The critical temperature and the condensate fraction in 
liquid helium}
 London had noticed that $T_{\lambda}$ was close to 
 the transition temperature of an ideal Bose gas with the same
 density. 
 \begin{equation}
     T_{BEC}= \left ( \frac{2\pi \hbar^{2}}{1.897 m k_{B}} \right ) 
     n^{2/3}
     \label{eq:BEC}
     \end{equation}
     
Inserting a number density $n\: =\: 2.18 \times 10^{22} \rm cm^{-3}$ in 
Eq.~\ref{eq:BEC} leads to $T_{BEC}$ = 3.1~K while $T_{\lambda}$ = 2.2~K.
London had also noticed that the molar volume of liquid helium
 was large, due to the large kinetic energy
 corresponding to the quantum fluctuations of atoms in the cage formed
 by their neighbours.  He had further noticed that the anomaly of the
 specific heat of liquid helium near $T_{\lambda}$ was not very
 different from the cusp one should expect in the case of the BEC of
 an ideal gas.  He thus claimed that , for the understanding of the 
 superfluid transition in helium,  ``it seems difficult not to imagine 
a connection with the condensation of the Bose-Einstein statistics''.
However, in the next sentence, he immediately added that ``a model 
which is so far away from reality that it simplifies liquid helium to an 
ideal gas'' cannot lead to the right value of its specific heat. 

In fact, the difficulty is much deeper than a question of specific 
heat only. As claimed by A.J. Leggett~\cite{Leggett}, 
it is not rigorously proven that a condensate, i.e. 
a macroscopic occupation of a ground state, should necessarily exist in a Bose 
liquid. It is not obvious either that there is a continuous path from the 
ideal Bose gas to the strongly interacting Bose liquid via the weakly 
interacting Bose gas where the existence of a condensation has been 
demonstrated both theoretically and , now, experimentally. 
The total absence of reference to BEC in Landau's work is obviously 
intentional. Landau deliberately ignored London~\cite{prize}. He 
must have considered that there was no continuity from the 
gas to the liquid. It is now generally accepted that London was 
right, but, as we shall see, this issue is quite subtle.

Let us keep a few 
important steps only in the long history of this controversy. In 1947, 
Bogoliubov~\cite{Bogoliubov} showed that, in a Bose gas 
with weak repulsive interactions, below the BEC transition, 
the low energy excitations are collective modes with a non-zero 
velocity. This is what allows such a weakly interacting Bose gas to 
be superfluid. Bogoliubov justified Landau's assumption that there 
are no single particle excitations at low energy, so that a 
superfluid current cannot dissipate its kinetic energy if its velocity 
is less than a critical value, here the velocity of this collective 
mode, which is identical to the ordinary sound velocity.

In my opinion, the next crucial step is the historical paper by O.~Penrose
and L.~Onsager~\cite{Penrose} who generalized BEC to an interacting 
system. The single particle states are no longer eigen states. 
They considered the one-body density matrix $\rho_{1}(r)$ and 
its eigenvalues in the limit where $r$ tends to infinity. They
correspond to the occupation of the various 
eigenstates of the whole liquid. If the occupation of the ground 
state is macroscopic, meaning that the ``condensate fraction'' $n_{0}$ is of 
order unity (when $r$ tends to infinity),
then  BEC has taken place.  Since the momentum distribution is the Fourier 
transform of 
this one-body density matrix, a finite $n_{0}$ means a delta 
function at $p$ = 0 in the momentum distribution: there is a ground state
with zero momentum 
which is occupied by a macroscopic fraction of the total number of 
particles. As explained by P. Sokol~\cite{Sokol}, 
the physical interpretation of $\rho(r)$ is the overlap 
of a wavefunction of the system when a particle is removed at the 
origin and replaced a distance $r$ away. The finite value of $n_{0}$ 
indicates a long range coherence which is often called ``ODLRO'' 
for ``Off diagonal long range order''. 
The corresponding eigen function is the ground state wave 
function which is the order parameter for the transition. It is 
remarkable to see that, in the same article, Penrose and Onsager 
used Feynmann's wave function for the condensate, approximated 
liquid helium as a gas of hard spheres, and ended with  $n_{0} 
\approx$ 8 \% at low pressure, in agreement with more recent results.
The ground state wave function has an amplitude and a phase $\phi$. 
The superfluid velocity is 
\begin{equation}
   v_{s} \: = \: \frac{ \hbar}{m} \nabla \phi
   \end{equation}

More theoretical efforts have been done in order to calculate $n_{0}$. 
In his review~\cite{Sokol}, Sokol cites several numerical methods from 
 ``Path integral Monte Carlo'' (PIMC) to ``Green's function Monte Carlo'' 
 (GFMC) which are more or less consistent with the rough estimate 
 first made by Penrose and Onsager. His GFMC predicts 9.2 \% 
 at low temperature and pressure. As for measuring
 the condensate fraction in liquid helium it has been another challenge. 
Contrary to what happens with trapped Bose gases, where the condensate 
is directly observed and easily measured, there is no experimental 
evidence for the existence of a condensate in liquid helium. The 
quantitative analysis  of ``Deep Inelastic Neutron scattering'' 
experiments is very delicate. Reliable experimental values for $n_{0}$ 
were obtained in the last decade only. They depend not only on the assumption 
that a condensate exists, but also on the particular shape of the 
distribution function for excited states with non-zero momentum. Given 
this ``caveat'', the presently accepted experimental value for $n_{0}$ in liquid 
helium at zero pressure and at low temperature is 10 $\pm$ 1.5 \% , in agreement with 
theory.

\begin{figure}
\centerline{\includegraphics[width=0.8\linewidth]{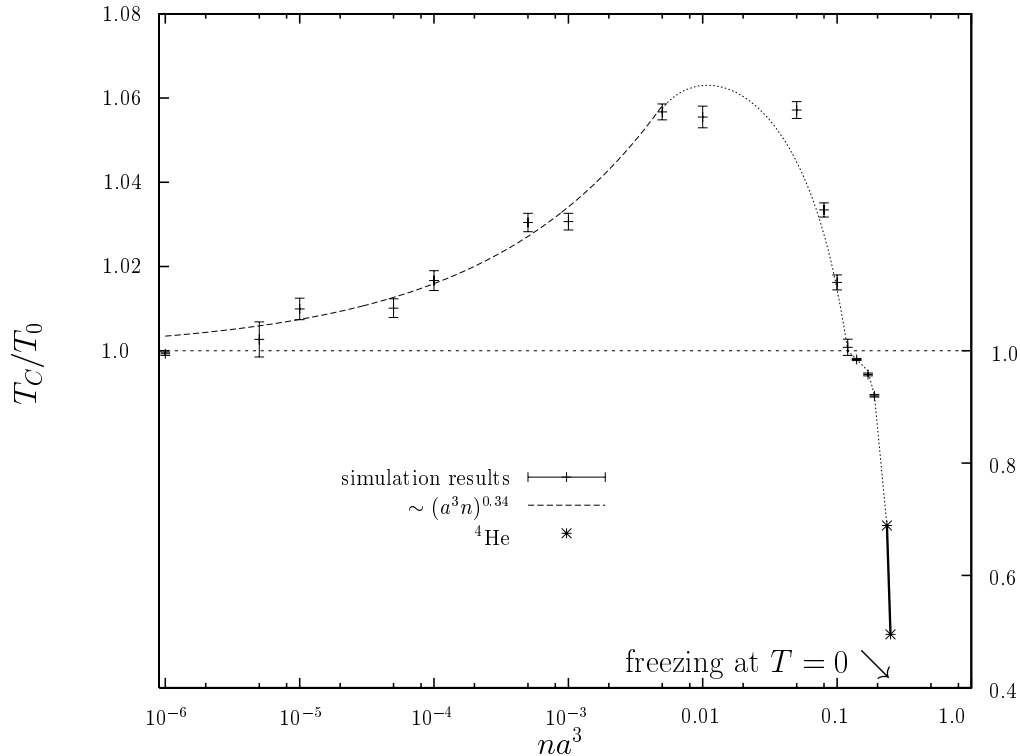}}
\caption{The calculation by Gruter et al. of the BEC transition
temperature in a hard sphere model as a function of the strength of 
the interactions. Note that the temperature scale is not the same for 
the low interaction part and for the large interaction part, so that 
an artificial cusp shows up where $T_{c}$ crosses the value $T_{0}$ of 
the ideal Bose gas. The star-like points on the right correspond to 
liquid helium. They lie on the same curve (dotted line) as for dilute 
gases.}
\label{fig:Gruter}
\end{figure}

It is now interesting to consider the pressure (or density) dependence 
of the condensate fraction. As shown by Sokol~\cite{Sokol} again, theory and 
experiment agree about this also:  $n_{0}$ decreases from about 10 \% at $P$ = 0 
bar where the density of liquid helium is $\rho$ = 0.145 $\rm 
g.cm^{-3}$ to about 5~\% only at the melting pressure $P_{m}$ = 25.3 bar 
where $\rho$ = 0.172 $\rm 
g.cm^{-3}$ (Remember that liquid helium does not solidify at low 
temperature except if a pressure greater than 25.3 bar is applied).
One understands the decrease of $n_{0}$ as a result of the exchange
between atoms becoming more 
difficult as the density increases in this range. A probably 
equivalent explanation is given by Bauer et al.~\cite{Bauer} who 
write that ``exchange is inhibited by the increase in effective mass 
of the particles''. This variation is consistent with the 
transition temperature $T_{\lambda}$ being also a decreasing function 
of pressure. One now realizes that this behaviour is opposite to the 
prediction from Eq.~\ref{eq:BEC}. Does it mean that interactions always 
decrease the critical temperature for condensation? Certainly not, as 
shown by Gruter, Ceperley and Laloe~\cite{Gruter}.
Gruter et al. calculated $T_{BEC}$ in a hard sphere fluid 
as a function of the parameter $na^{3}$ which describes the 
interaction strength ($n$ is the number density and $a$ is either a 
scattering length in the dilute case or a hard core in the dense 
limit of liquid helium). They found that the variation of $T_{BEC}$ 
is non-monotonic, a quite remarkable result. At low density, some 
interactions help the system to be homogeneous and to establish a long range 
coherence. On the contrary, large interactions make 
exchange difficult at large density. On  Fig.~\ref{fig:Gruter}, 
one sees that they were able to put the transition temperature 
of liquid helium on the same curve as their prediction for the 
critical temperature of BEC in dilute Bose gases. This curve would 
have surprised Landau !

\begin{figure}
\centerline{\includegraphics[height=0.5\linewidth]{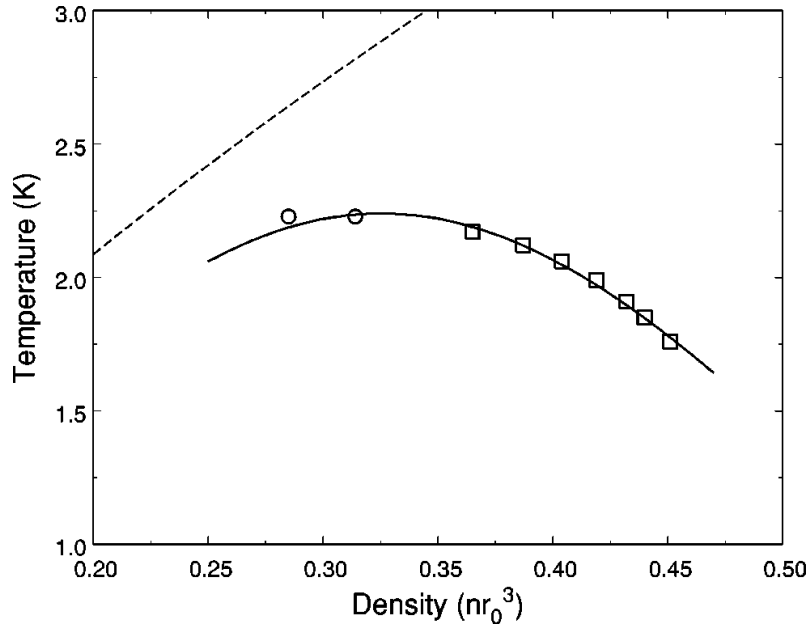}}
    \caption{The density variation of the lambda transition according 
    to Apenko (solid line and symbols).  The lambda temperature approaches
    the BEC temperature of the ideal Bose gas (dotted line) 
    in the metastable region of liquid helium at low density.
     This behavior is consistent with the results 
    obtained by Bauer et al.}
    \label{fig:Apenko}
\end{figure}

I would like to make two further remarks on the pressure variation of 
$T_{\lambda}$ in liquid helium. Suppose now that, by some clever 
mean, one could study liquid helium at a density lower than its 
equilibrium value. Would $T_{\lambda}$ come closer to the ideal gas 
value $T_{BEC}$? 
We have shown that this is possible by using acoustic pulses of high 
intensity and short duration~\cite{Balibar02}. In the absence of walls 
and impurities, it is possible to drive liquid helium to a metastable 
state at a negative pressure which approaches the ``spinodal limit'' 
at - 9.5~bar. Under such conditions, the density of liquid helium can 
be lowered to 0.10 $\rm g.cm^{3}$, about 30 \% less than in 
equilibrium at the saturated vapor pressure. Apenko~\cite{Apenko} and 
Bauer~\cite{Bauer} have calculated $T_{\lambda}$ in this metastable 
region of the phase diagram (see Fig.~\ref{fig:Apenko}). They both
predicted that $T_{\lambda}$ should reach 
a maximum value of about 2.2~K for a density of 0.12 $\rm g.cm^{3}$ 
corresponding to -8~bar.  In summary, in 
liquid helium at negative pressure, $T_{\lambda}$ ceases to be a 
decreasing function of pressure and comes closer to the ideal gas 
behaviour. This is consistent with our observation of a cusp in the 
cavitation pressure at 2.2~K~\cite{Balibar02}. We naturally
attributed this cusp to 
the crossing of the lambda transition at negative pressure, and we 
found it where it is predicted by recent theories, not on a linear 
extrapolation of the behavior at positive pressure. It is the sign 
that rotons are no longer dominating the thermodynamics of liquid 
helium near its spinodal limit. 

We are also using the same acoustic technique to study overpressurized 
liquid helium. Our main motivation is to look for a possible limit of
instability for the metastable liquid at high pressure with respect 
to the formation of the crystalline phase which is the stable one. In 
the course of this search, we have already reached about 120 bar 
without seing nucleation of helium crystals. This shows that it is 
possible to study liquid helium at much higher densities than 
previously thought. In his review , Sokol ~\cite{Sokol} considers 
this region as ``unaccessible'' but he was not aware of our experiments. 
It is thus not absurd to consider what happens to superfluidity in 
liquid helium at high density. A crude extrapolation of Sokol's 
calculation of $n_{0}$ would predict that it becomes vanishingly small 
near a density of about 0.19 $\rm g.cm^{-3}$. Given the known equation 
of state~\cite{Balibar02}, it would correspond to a pressure of order 
55~bar, i.e. 30~bar above the equilibrium melting pressure. We have 
driven liquid helium up to a much higher density. It is 
hard to imagine that, if $n_{0}$ is very small $T_{\lambda}$ is not 
very small as well. How exactly does the lambda line extrapolate in 
this metastable region? Is it possible that a helium glass exists ? It 
is certainly not easy to study such speculative properties but we 
might obtain some information on it from our experiments. It would 
also be interesting to calculate the properties of a quantum hard sphere 
model in the very high density limit where the system jams, not only 
in the very dilute case where the system approaches the ideal gas.

With these few comments and ideas, I hope that I have provided some 
information for a comparison between superfluid liquids and 
superfluid gases. My purpose was also to show that a few questions 
deserve further study.

\end{document}